\documentclass[aps,prl,reprint,groupedaddress,showpacs]{revtex4-1}
\usepackage{amsmath,amssymb}
\usepackage{graphicx}

\begin{document}

\title{Magnetic-Field-Induced Superconductivity in Ultrathin Pb Films\\
with Magnetic Impurities}

\author{Masato Niwata, Ryuichi Masutomi, and Tohru Okamoto}
\affiliation{Department of Physics, University of Tokyo, 7-3-1 Hongo, Bunkyo-ku, Tokyo 113-0033, Japan}

\date{October 23, 2017}

\begin{abstract}
It is well known that external magnetic fields 
and magnetic moments of impurities both suppress superconductivity.
Here, we demonstrate that their combined effect enhances
the superconductivity of a few atomic layer thick Pb films grown on a cleaved GaAs(110) surface.
A Ce-doped film, where superconductivity is totally suppressed at a zero field,
actually turns superconducting
when an external magnetic field is applied parallel to the conducting plane.
For films with Mn adatoms, the screening of the magnetic moment
by conduction electrons, i.e., the Kondo singlet formation, becomes important.
We found that the degree of screening can be reduced by capping the Pb film with a Au layer,
and observed the positive magnetic field dependence of the superconducting transition temperature.
\end{abstract}


\maketitle

In 1914, Kamerlingh Onnes reported the destructive effect of a magnetic field 
on the superconductivity of Pb and Sn \cite{Onnes1914}.
Today, it is conventional wisdom that external magnetic fields, as well as magnetic impurities, 
break the time-reversal symmetry of Cooper pairs and tend to suppress superconductivity.
There are two mechanisms responsible for this: the orbital effect (OE) 
and the paramagnetic effect (PE) \cite{Clogston1962,Chandraskhar1962}. 
In the case of a superconductor containing magnetic elements, 
localized magnetic moments are also affected by the magnetic field. 
This should lead to an additional effect on superconductivity 
while it is usually obscured by OE and/or PE. 
For atomically thin or layered materials, OE can be eliminated 
by setting the magnetic field direction parallel to the conducting plane. 
Recently, it has been demonstrated that
the reduction of the superconducting transition temperature $T_c$
due to the parallel magnetic field $H_\parallel$ is extremely small ($\sim 1\%$ for 10~T)
in ultrathin Pb films grown on a cleaved GaAs(110) surface \cite{Sekihara2013}. 
The suppression of PE was explained in terms of the spin precession 
due to the Rashba field \cite{Rashba1960,Bychkov1984}, which allows nonmagnetic scattering 
by defects to mix the spin-up state and spin-down state \cite{Sekihara2015}. 
In this study, we employ ultrathin Pb films as the host superconductor to minimize OE and PE, 
which act directly on conduction electrons, and investigate the magnetic-field effect 
on superconductivity related to magnetic impurities. 
For films with Ce, we observe pronounced positive $H_\parallel$ dependence of $T_c$.
Furthermore, in a Pb-Ce(10 at\%) alloy film, $H_\parallel$ actually induces
a quantum phase transition from the normal state to the superconducting state.
These results are consistent with a recent theory by Kharitonov and Feigelman
\cite{Kharitonov2005}.
In the case of deposition of $3d$ transition metals, 
the exchange coupling is strong and
the localized moment of an adatom is expected to be screened by conduction electrons,
forming the Kondo singlet state \cite{Wilson1975}.
We find that the exchange coupling is weakened on a Pb film covered with a Au layer
and observe the positive $H_\parallel$ dependence of $T_c$ for Mn deposition.

The films were grown by vapor deposition
onto a nondoped insulating GaAs single-crystal substrate,
which was cooled down to liquid helium temperatures
to avoid grain formation and impurity segregation.
In the case of pure Pb,
superconductivity can be observed even in the monolayer regime
due to an atomically flat surface of the cleaved GaAs substrate
\cite{Sekihara2013}.
Current and voltage electrodes were prepared in advance
by the deposition of gold films onto noncleaved surfaces \cite{Okamoto2011,Masutomi2015}.
The cleavage of GaAs, the deposition of metals,
and resistance measurements were performed {\it in situ} under ultrahigh vacuum conditions.
Since the superconducting transition temperature $T_c$ did not change
when the film was left overnight,
we believe that the base pressure was low enough and contamination effects were negligible.
The amount deposited was measured with a quartz crystal microbalance
and determined with an accuracy of about 5\%.
The four-probe resistance of the ultrathin film 
on a cleaved GaAs(110) surface (4 $\times$ 0.35 mm${}^2$)
was measured using the standard lock-in technique at 13.3 Hz.
The magnetic-field direction with respect to the surface normal
was precisely controlled using a rotatory stage
on which the sample was mounted, together with a Hall generator,
a RuO$_2$ resistance thermometer, and a heater.
The sample stage can be cooled to 0.5~K via a silver foil
linked to a pumped ${}^3$He refrigerator.
All the data were taken when the temperature of the sample stage
was kept constant so as to ensure thermal equilibrium
between the sample and the thermometer.
The magnetoresistance effect of the RuO$_2$ resistance thermometer
was systematically calibrated against the vapor pressure of
the liquid ${}^3$He or ${}^4$He for various temperatures \cite{RuO2}.
After the correction, $T_c$ can be
determined with a relative accuracy of better than 0.2\%.

\begin{figure*}[t!]
\begin{center}
\includegraphics[width=1.0\linewidth, clip]{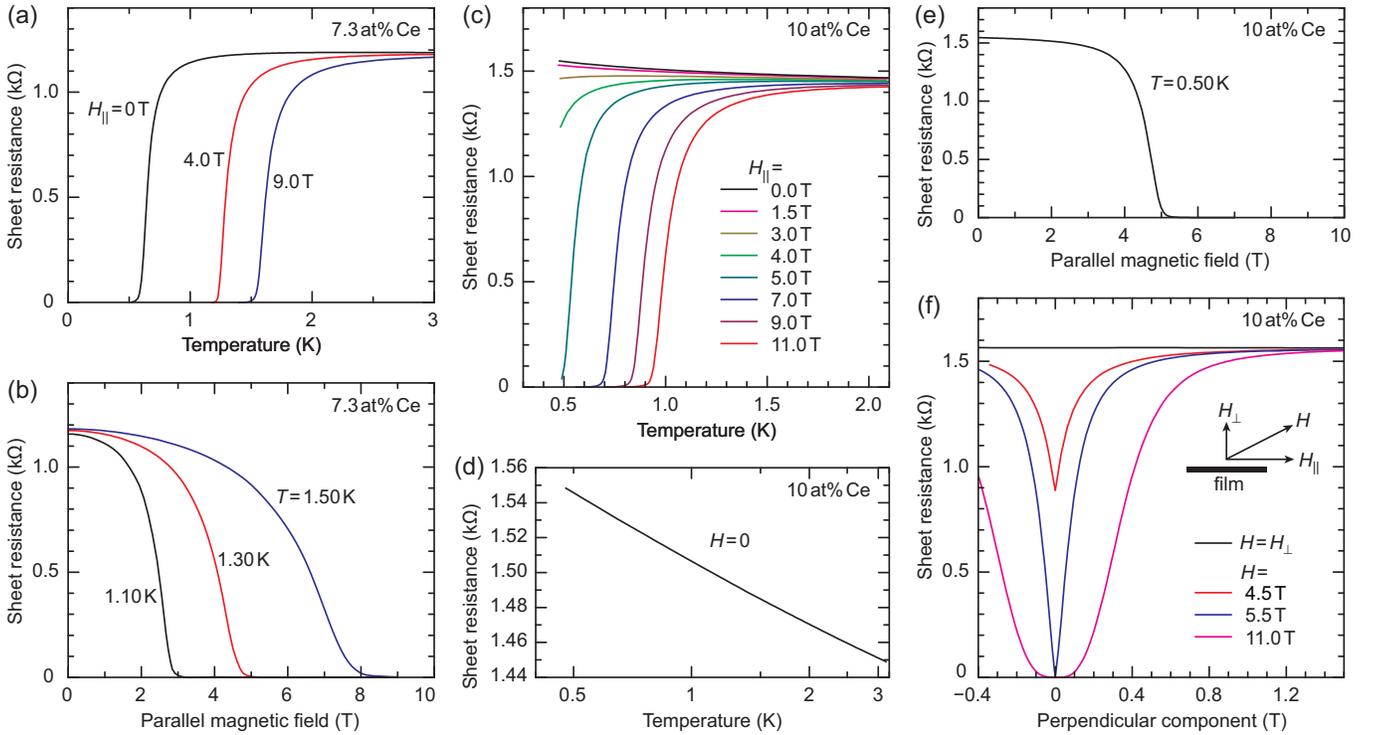}
\caption{
Superconductivity of Pb-Ce alloy films
enhanced or induced by parallel magnetic fields.
(a) Sheet resistance $R_{\rm sq}$ of $F1$ (7.3~at\% Ce)
versus $T$ for different $H_\parallel$.
(b) $H_\parallel$ dependence of $R_{\rm sq}$ of $F1$ at different temperatures.
(c) $R_{\rm sq}$ of $F2$ (10.0~at\% Ce) versus $T$ for different $H_\parallel$.
(d) Expanded plot of the zero-magnetic-field data in (c).
(e) $H_\parallel$ dependence of $R_{\rm sq}$ of $F2$ at $T=0.50$~K.
(f) $H_\perp$ dependence of $R_{\rm sq}$ of $F2$ at $T=0.50$~K
for different values of $H=(H_\perp^2+H_\parallel^2)^{1/2}$.
}
\end{center}
\end{figure*}

First, we studied the superconducting transition
of a Pb film with an average thickness of $d=0.65$~nm
for different amounts of Ce deposition.
While $T_c$ decreases monotonically with increasing Ce coverage \cite{CeDepo},
the suppression rate is small.
Even for a coverage near one monolayer, 
$T_c$ decreases only by 40\% (from 3.2 to 1.9~K).
To enhance the exchange coupling between the localized moment of Ce and conduction electrons,
we prepared two Pb-Ce alloy films ($F1$ and $F2$) where Ce atoms
are expected to be dominantly surrounded by Pb atoms.
They are 1.1~nm thick (approximately 3 atomic layers)
and were formed from threefold alternate depositions of Ce and Pb.
Figure 1(a) shows the $T$ dependence of the sheet resistance $R_{\rm sq}$
of $F1$ (containing 7.3 at\% Ce) for different values of $H_\parallel$.
At zero magnetic field, superconductivity is strongly suppressed and
$R_{\rm sq}$ becomes zero only below 0.6~K.
With increasing $H_\parallel$, however, the superconducting transition obviously shifts to higher temperatures.
In Fig. 1(b), $R_{\rm sq}$ is plotted as a function of $H_\parallel$ for fixed temperatures
at which the $H_\parallel$-induced resistance drop is clearly observed.

Figure 1(c) shows the $T$ dependence of $R_{\rm sq}$ of $F2$ (containing 10 at\% Ce)
with varying $H_\parallel$.
At zero magnetic field,  $\partial R_{\rm sq}/\partial T$ is negative and
no superconducting behavior is observed down to 0.47~K [see also Fig. 1(d)].
This film becomes superconducting only in high parallel magnetic fields.
In Fig. 1(e), the $H_\parallel$-induced resistance drop at $T=0.50$~K is shown.
In the presence of the perpendicular magnetic field component $H_\perp$,
OE destroys the superconducting state.
Figure 1(f) shows the $H_\perp$ dependence of $R_{\rm sq}$ of $F2$ at $T=0.50$~K
obtained for different total strength $H$.
For large $H_\perp$, $R_{\rm sq}$ approaches the normal-state value 
irrespective of $H$.
The superconducting region around $H_\perp =0$ becomes wider as $H$ increases from 5.5 to 11~T,
indicating the stabilization of superconductivity with respect to OE,
as well as with respect to thermal fluctuations.

Magnetic-field-induced superconductivity has been observed before in several other materials
\cite{Meul1984,Lin1985,Uji2001,Konoike2004,Levy2005}.
Except for the special case of a spin-triplet superconductor URhGe \cite{Levy2005}, 
it was understood in terms of the Jaccarino-Peter (JP) mechanism \cite{Jaccarino1962},
where PE of a mean field produced by aligned local magnetic moments
through the antiferromagnetic exchange interaction is compensated by the external magnetic field. 
However, this mechanism is unlikely to account for our results.
As described above, PE is strongly suppressed in ultrathin Pb films,
for which the Pauli-limiting field is estimated to be on the order of 100~T.
Furthermore, since the mean field is absent at zero magnetic field,
where the magnetic moments are expected to be randomly oriented,
the JP model does not explain why 
superconductivity is suppressed at $H=0$.

The enhancement of $T_c$ by $H_\parallel$ in Pb films without intentional impurities
was reported in Ref.~[\onlinecite{Gardner2011}].
The maximum enhancement reaches 13.5\% at $H_\parallel =8$~T for $d=1.2$~nm.
However, no theory adequately explains 
the positive $H_\parallel$ dependence of $T_c$ in pure Pb.
For Pb films grown on a cleaved GaAs surface,
we did not observe the enhancement of $T_c$ by $H_\parallel$
at least in the range $0.22 \leq d \leq 3.0$~nm
unless magnetic impurities were added.
This discrepancy might be due to the difference in substrates and film morphology.

According to the pioneering theory of Abrikosov and Gor'kov \cite{Abrikosov1961},
the exchange scattering of electrons by magnetic impurities suppresses superconductivity.
Recently, Kharitonov and Feigelman (KF) have shown that
the pair-breaking effect is weakened by the polarization of impurities
in a magnetic field \cite{Kharitonov2005}.
While the polarization increases the rate of scattering without spin flip,
it decreases the spin-flip scattering rate.
The total exchange scattering rate is reduced from its zero-magnetic-field value $\nu_s$
to $\nu_s S/(S+1)$ as the magnetic field increases to infinity.
Here, $S$ is the spin of the impurity.
It has been predicted that a magnetic field can induce a superconducting quantum phase transition
when $\nu_s$ is in an appropriate range
and PE is strongly reduced due to spin-orbit scattering.
In ultrathin Pb films, high spin-orbit scattering rate is expected to be effectively achieved
by the combination of the Rashba field and nonmagnetic scattering \cite{Sekihara2015}.
In Fig. 2(a), $T_c$ in $F1$ and $F2$,
which is defined as the temperature at which $R_{\rm sq}$ is half the normal-state value $R_N$,
is plotted as a function of $H_\parallel$.
Solid curves are the best fits based on the KF theory, assuming $S=1/2$ for Ce and
neglecting OE and PE.
\begin{figure}[b]
\begin{center}
\includegraphics[width=0.75\linewidth, clip]{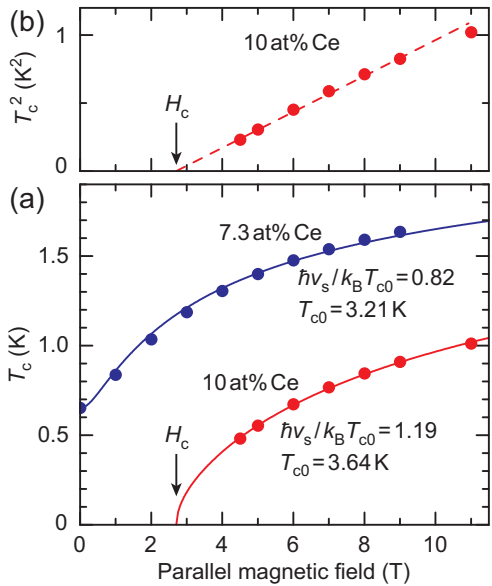}
\caption{
Field-temperature phase diagram for Pb-Ce alloy films.
(a) Circles are the $H_\parallel$ dependence of $T_c$ for $F1$ (blue) and $F2$ (red).
Solid curves are calculations according to Ref.~[\onlinecite{Kharitonov2005}] with $S=1/2$.
(b) Square of $T_c$ for $F2$.
}
\end{center}
\end{figure}
The calculation well reproduces the experimental results with only two fitting parameters,
$\nu_s$ and $T_{c0}$ ($T_c$ in the absence of exchange scattering).
The ratio $\hbar \nu_s/k_B T_{c0}$ exceeds a critical value of 0.882 \cite{Kharitonov2005}
for $F2$, while it does not for $F1$.
The quantum phase transition is calculated to occur at $H_c =2.7$~T.
Near the critical point, $T_c$ varies approximately as $(H_\parallel-H_c)^{1/2}$.
To see this, $T_c^2$ is plotted for $F2$ in Fig. 2(b).
The experimental data almost fall on a straight line, which cuts the $H_\parallel$-axis at $H_c$.
According to the present calculation,
$T_c$ increases monotonically with $H_\parallel$ and approaches a constant value.
This is in contrast to the JP mechanism, where the superconductivity 
is suppressed again when the magnetic field increases further.

While the exchange scattering in the Pb-Ce alloy films strongly suppresses superconductivity,
it has almost no effect on the normal-state resistance.
The exchange scattering time $\tau_s =\nu_s^{-1}$ (1.8~ps for $F2$)
is 3 orders of magnitude longer than
the transport scattering time $\tau_t$,
which can be estimated from $R_N$.
Therefore, the Kondo effect \cite{Kondo1964}
cannot be the main cause of the logarithmic $T$ dependence shown in Fig. 1(d),
which may be attributed to the weak localization effect in disordered two-dimensional systems
\cite{Lee1985}.

The exchange coupling between a localized moment and conduction electrons is 
much stronger for $3d$ impurities than for $4f$ impurities.
Although a small amount of Mn or Cr strongly suppresses
the superconductivity of ultrathin Pb films,
we did not observe the enhancement of $T_c$ by $H_\parallel$
when we deposited Mn or Cr directly onto the Pb film.
This can be attributed to the formation of the Kondo singlet state \cite{Wilson1975},
which was not taken into account in the KF theory.
The Kondo temperature $T_K$, below which this state exists,
depends exponentially on the exchange coupling constant $J$
and can have a wide range of values.
For $T_K \gg T_{c0}$, the pair-breaking effect of the localized moment vanishes
and instead $T_c$ is suppressed by an effective repulsive interaction between 
Cooper-pair electrons
through the virtual polarization of the Kondo singlet state
\cite{Matsuura1977,Sakurai1978}.
In this regime, the suppression rate of $T_c$ with magnetic impurities increases
as $T_K$ decreases.
For  $T_K \ll T_{c0}$, on the other hand,
$T_c$ is suppressed by the unscreened moment
and the suppression rate is proportional to $J^2$.
We believe that the Pb-Ce films are in this regime
since the $H_\parallel$ dependence of $T_c$ is successfully explained in terms of the KF theory.
As schematically illustrated in Fig. 3(a),
the suppression rate is expected to have a maximum as a function of $T_K$ or $|J|$ \cite{Matsuura1977}.

\begin{figure}[t]
\begin{center}
\includegraphics[width=1.0\linewidth, clip]{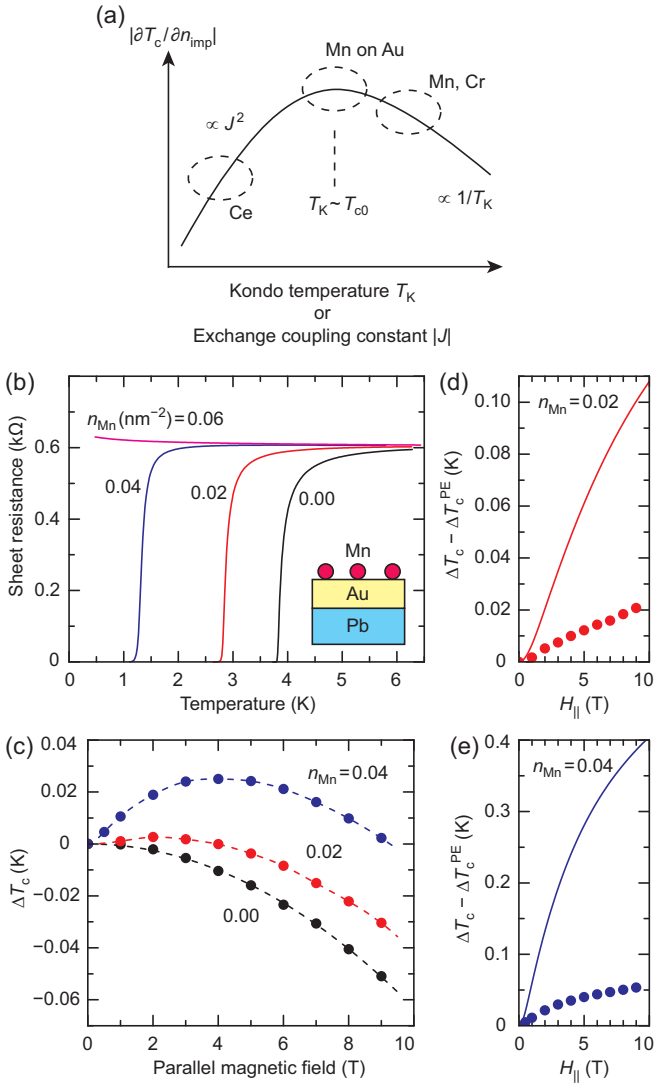}
\caption{
Superconductivity of a Au/Pb film with Mn adatoms.
(a) Schematic dependence of the suppression rate of $T_c$ per magnetic impurity concentration
on $T_K$ (or $|J|$).
(b) $T$ dependence of $R_{\rm sq}$ of a Au/Pb film at $H=0$ with varying $n_{\rm Mn}$.
(c) 
$H_\parallel$-induced change in $T_c$ ($\Delta T_c$) for different $n_{\rm Mn}$.
(d), (e) $H_\parallel$ dependence of $\Delta T_c$ after subtracting PE.
Solid curves are calculations based on Ref.~[\onlinecite{Kharitonov2005}].
}
\end{center}
\end{figure}

To reduce the exchange coupling between the localized moment and conduction electrons,
a 0.3~nm layer of Au was used to cover an ultrathin Pb film ($d=1.1$~nm) before Mn deposition.
Figure 3(b) shows the $T$ dependence of $R_{\rm sq}$ at $H=0$
for different densities $n_{\rm Mn}$ of Mn adatoms.
The suppression rate is 2 orders of magnitude greater than that for Ce deposition \cite{CeDepo}
and superconductivity is fully suppressed for $n_{\rm Mn}=0.06~{\rm nm}^{-2}$ ($\approx 0.006$ monolayer).
In Fig. 3(c), the $H_\parallel$-induced change in $T_c$ is shown.
Besides the negative quadratic $H_\parallel$ dependence arising from PE,
a small enhancement of $T_c$ is observed after Mn deposition.
Since PE is expected to depend only on $\tau_t$ for strong Rashba splitting 
\cite{PE},
it can be estimated from the data for $n_{\rm Mn}=0$.
After subtracting PE, the results for $n_{\rm Mn}=0.02$ and $0.04~{\rm nm}^{-2}$
are plotted in Figs. 3(d) and 3(e), respectively.
The solid curves are calculated based on the KF theory
using $S=5/2$ and $T_{c0}=3.90$~K obtained from $T_c$ for $n_{\rm Mn}=0$.
The experimental enhancement of $T_c$ is several times smaller than the calculation.
This seems reasonable if the film is in the regime of $T_K \sim T_{c0}$,
where the localized moment is partially screened by conduction electrons.
In recent years, there has been a great interest in
the Kondo screening of surface-adsorbed magnetic atoms and molecules
\cite{Li1998,Madhavan1998,Bogani2008,Franke2011,Zhang2015,Huang2017},
partly due to their potential applications in nanoscale spintronic devices.
It has been demonstrated here that the study of
the $H_\parallel$ dependence of $T_c$ can provide valuable information 
on the degree of Kondo screening.

In conclusion, we observed a $H_\parallel$-induced
superconducting quantum phase transition
in a Pb-Ce(10 at\%) alloy film.
The $H_\parallel$ dependence of $T_c$ is 
well reproduced by the calculation based on
a recent theory by Kharitonov and Feigelman \cite{Kharitonov2005},
who considered the reduction of the exchange scattering rate in magnetic fields.
This is in contrast to previously reported magnetic-field-induced superconductors
\cite{Meul1984,Lin1985,Uji2001,Konoike2004},
which are attributed to the Jaccarino-Peter mechanism \cite{Jaccarino1962}.
While the formation of the Kondo singlet state \cite{Wilson1975} is not taken into account in the KF theory,
it becomes important for 3$d$ impurities.
We found that the degree of Kondo screening of Mn adatoms
can be reduced by capping the Pb film with a Au layer,
and observed the positive $H_\parallel$ dependence of $T_c$.

\begin{acknowledgments}
This work was supported by JSPS KAKENHI Grants No. JP26287072 and No. JP16H03998.
\end{acknowledgments}

\clearpage

\section{Supplemental material}

\subsection{Magnetoresistance effect of the RuO$\bm{{}_2}$ resistance thermometer}

The RuO$_2$ resistance thermometer (RO-600) was manufactured by Scientific Instruments (SI), Inc.,
which offers a simple linear formula for a positive magnetoresistance.
It allows the correction of any apparent temperature,
measured in a magnetic field of up to 16~T, to the actual temperature
with an error of less than $\pm 1.6$\%
for a broad temperature range of 0.036 to 4.2~K.
To obtain better accuracy, 
we performed a systematic calibration up to 14~T
using the vapor pressure of the liquid ${}^3$He or ${}^4$He.
Our measurements reveal small $T$ dependence
while the variation is almost within the error bars for the SI's formula.

\begin{figure}[b]
\begin{center}
\includegraphics[width=0.9\linewidth, clip]{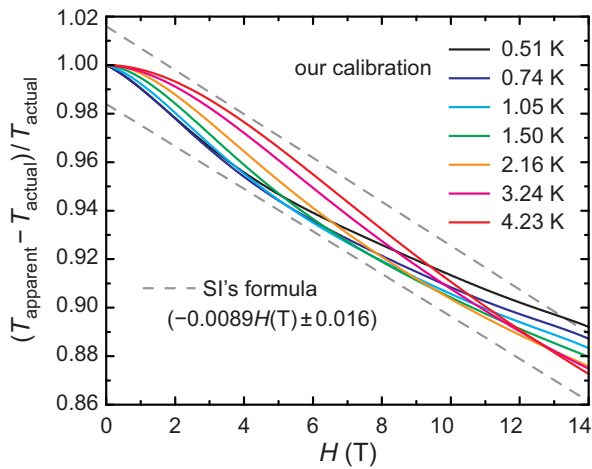}
\caption{
Deviation of the apparent temperature indicated by a RO-600 thermometer
from the actual temperature. 
}
\end{center}
\end{figure}

\subsection{Pb film covered with Ce adatoms}

As mentioned in the main text,
we have performed transport measurements 
also on a 0.65~nm thick Pb film covered with Ce adatoms.
Figure 5 shows the $T$ dependence of the sheet resistance $R_{\rm sq}$
at $H=0$  for different densities of Ce adatoms,
$n_{\rm Ce}=0, 1.9, 3.7, 5.6, 7.4$~nm${}^{-2}$.
Since $n_{\rm Ce}=7.4$~nm${}^{-2}$ is comparable with
the planar densities of the bulk Ce(111) plane (8.7 nm${}^{-2}$) and
the bulk Pb(111) plane (9.4 nm${}^{-2}$),
it corresponds to a coverage near one monolayer.
The superconducting transition shifts to lower temperatures
with increasing $n_{\rm Ce}$.
As in the case of the Pb-Ce alloy films,
it can be attributed to the exchange scattering of conduction electrons.
In Fig. 6, $T_c$ is shown as a function of $H_\parallel$.
The enhancement of $T_c$ by $H_\parallel$ is observed after Ce deposition.
It increases with increasing $n_{\rm Ce}$
while it is smaller than that observed in the Pb-Ce alloy films (see Fig. 2(a) of the main text).

\begin{figure}[t]
\begin{center}
\includegraphics[width=0.8\linewidth, clip]{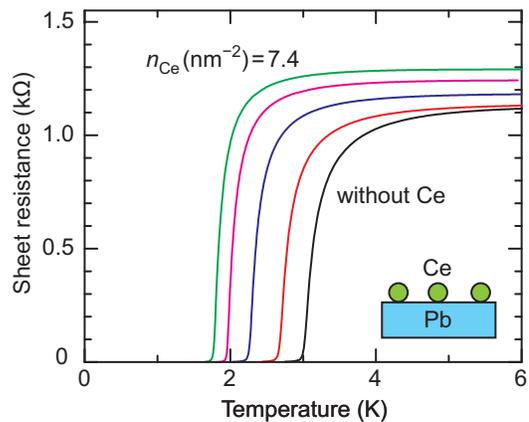}
\caption{
Temperature dependence of the sheet resistance of
a 0.65~nm thick Pb film at $H=0$
for different Ce coverages.
}
\end{center}
\end{figure}
\begin{figure}[t]
\begin{center}
\includegraphics[width=0.8\linewidth, clip]{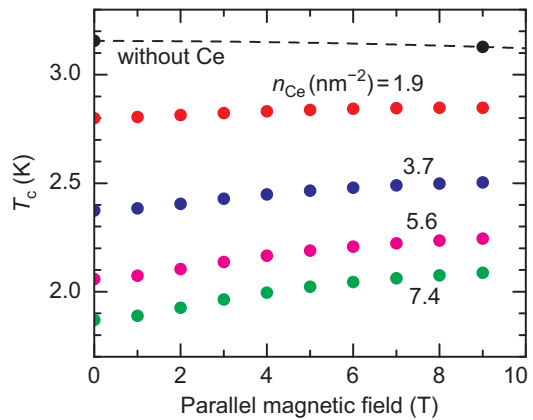}
\caption{
Parallel magnetic field dependence of the superconducting transition temperature of
a 0.65~nm thick Pb film for different Ce coverages.
}
\end{center}
\end{figure}
\begin{figure*}[t!]
\begin{center}
\includegraphics[width=0.4\linewidth, clip]{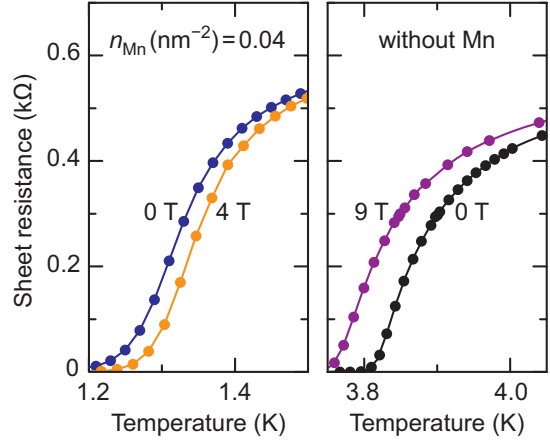}
\caption{
Parallel-magnetic-field-induced shift in the superconducting transition in
a Au/Pb film with and without Mn adatoms.
}
\end{center}
\end{figure*}
\begin{figure*}[t!]
\begin{center}
\includegraphics[width=0.8\linewidth, clip]{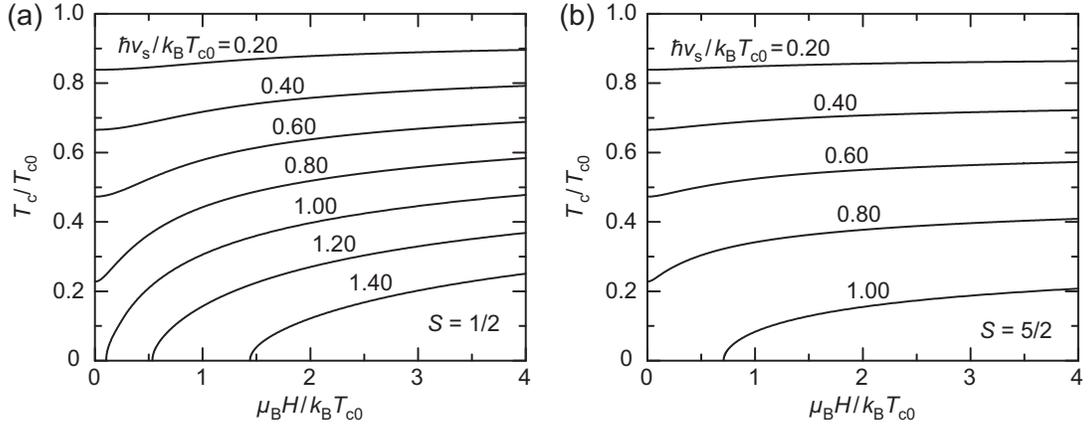}
\caption{
\textit{H-T} phase diagram based on the theory by Kharitonov and Feigelman.
Results for different exchange scattering rates are shown for the case without OE and PE.
(a) $S=1/2$. (b) $S=5/2$.
}
\end{center}
\end{figure*}

\subsection{Au/Pb film with Mn adatoms}

The $T$ dependence of $R_{\rm sq}$ of a Au/Pb film with Mn adatoms at $H=0$
has been shown in Fig. 3(b) of the main text.
In Fig. 7, $H_\parallel$ induced shifts are presented.
Before Mn deposition, the superconducting transition shifts to lower temperature
with increasing $H_\parallel$.
After Mn deposition, on the other hand, it shifts to higher temperature 
as $H_\parallel$ increases from 0 to 4~T.
As described in the main text,
the data were carefully taken under thermal equilibrium
with a RuO${}_2$ resistance thermometer calibrated in magnetic fields.
While $T_c$ was defined as the temperature
at which $R_{\rm sq}$ becomes $0.5 R_N$,
the $H_\parallel$ induced change $\Delta T_c$ does not significantly depend on
the criterion resistance value.
The $H_\parallel$ dependence of $\Delta T_c$ was shown in Fig. 3(c) of the main text.

\subsection{Calculations based on the theory by Kharitonov and Feigelman}

In the presence of the exchange scattering,
$T_c$ at $H=0$ is determined from the equation
\begin{equation}
\ln \left( \frac{T_c}{T_{c0}} \right)+\psi \left( \frac{1}{2} +
\frac{\hbar \nu_s}{2 \pi  k_B T_c}
\right) - \psi \left( \frac{1}{2}\right) =0.
\nonumber\\
\end{equation}
\noindent
Here, $T_{c0}$ is the transition temperature in the absence of exchange scattering,
$\psi (x)$ is the digamma function,
and $\nu_s$ is the exchange scattering rate at $H=0$.
This equation is equivalent to Eq. (1) in Ref.~[\onlinecite{Kharitonov2005}].
As $\nu_s$ increases, $T_c$ monotonically decreases from $T_{c0}$
and becomes zero for the critical scattering rate 
$\nu_s^\ast = \pi T_{c0}/2 e^\gamma  = 0.882 T_{c0}$,
where $\gamma=0.577$ is the Euler's constant.
The scattering rate can be represented as 
the sum of the spin-flip scattering rate and 
the rate of scattering without spin flip.
When $H$ increases from zero to infinity,
the former decreases from $2 \nu_s /3$ to zero 
while the latter increases from  $\nu_s /3$ to $\nu_s S/(S+1)$.
Thus, the total scattering rate decreases from $\nu_s$ to $\nu_s S/(S+1)$.
For $\nu_s^\ast < \nu_s < \nu_s^\ast (S+1)/S$,
if OE and PE are absent,
$H$ induces a superconducting quantum phase transition.

In finite magnetic fields, $T_c$ is calculated numerically using
Eqs. (4)-(6) and (9) in Ref. 1.
The results for different $S$ and $\nu_s$ are shown in Fig. 8.
The $H$ dependence of $T_c$ is stronger for $S=1/2$ (Ce) than for $S=5/2$ (Mn).

\begin{thebibliography}{99}
\bibitem{PE}
The coefficient for the negative quadratic $H_\parallel$ dependence of $T_c$ arising from PE
is proportional to the effective spin-orbit scattering time, which can be replaced by $\tau_t$
when the spin precession frequency in the Rashba field is high enough \cite{Sekihara2015}.
\end{thebibliography}
\end{document}